\documentclass[12pt]{article}                                                
\usepackage{amsmath,amssymb,amsfonts,amsbsy}                                 
\usepackage{cite}                                                            
\usepackage{graphicx}                                                        
\usepackage{wrapfig}                                                         
\usepackage{subfig}                                                          
\usepackage{epsfig}                                                          
                                                                             
\usepackage{bbm} 
\usepackage{bm} 
\usepackage{multicol}                                                        
\usepackage{dsfont} 
\usepackage{latexsym} 
\usepackage{lscape} 
\usepackage{mathrsfs} 
\usepackage{morefloats} 
\usepackage{floatflt} 
\usepackage{slashed} 
\usepackage{psfrag}
\usepackage{color}
\definecolor{darkgreen}{rgb}{0,0.65,0}

\DeclareFontFamily{OT1}{mygreek}{}%
\DeclareFontShape{OT1}{mygreek}{m}{n}{<->omsegr}{}%
\DeclareFontShape{OT1}{mygreek}{b}{n}{<->omsegrb}{}%
\DeclareFontShape{OT1}{mygreek}{m}{it}{<->omsegri}{}%
\DeclareFontShape{OT1}{mygreek}{bx}{n}{<->sub * mygreek/b/n}{}%
\DeclareFontShape{OT1}{mygreek}{m}{sl}{<->sub * mygreek/m/it}{}%
\DeclareSymbolFont{Greekrm}{OT1}{mygreek}{m}{n}                              
\DeclareSymbolFont{Greekbf}{OT1}{mygreek}{b}{n}                              
\DeclareSymbolFont{Greekit}{OT1}{mygreek}{m}{it}                             
\DeclareMathSymbol{\omegab}{\mathalpha}{Greekbf}{119}                        
                                                                             
\begin{document}                                                             
\addcontentsline{toc}{subsection}{{An analytic review of DSPIN-11}\\
{\it A.V. Efremov and J. Soffer}}

\setcounter{section}{0} \setcounter{subsection}{0} 
\setcounter{equation}{0} \setcounter{figure}{0} 
\setcounter{footnote}{0} \setcounter{table}{0}

\begin{center}
\textbf{AN ANALYTIC REVIEW OF DSPIN-11}

\vspace{5mm}

 {A.V. Efremov}${^1}$ and J. Soffer${^2}$  
\vspace{5mm}

\begin{small}
 $^1$\emph{Joint Institute for Nuclear Research, Dubna, Russia} \\
  \emph{E-mail: efremov@theor.jinr.ru}\\
 $^2$\emph{Department of Physics, Temple University 
   Philadelphia, Pennsylvania 19122-6082, USA}\\
   \emph{E-mail: jacques.soffer@gmail.com}
\end{small}
\end{center}

\vspace{0.0mm} 

\begin{abstract}
A short analytical review of the main results of the DSPIN-11 Workshop 
(JINR, Dubna, September 20--24, 2011) is given. 
\end{abstract}

\vspace{7.2mm} 

The XIV-Workshop on high energy spin physics (DSPIN-11) continued a 
series of similar conferences, the first of which took place 30 
years ago in 1981 on the initiative of the outstanding 
theoretical physicist L.I.Lapidus. Since then each odd year (in 
even years large International Symposia in spin physics take 
place) similar conferences have been organized in Protvino or in 
Dubna. They give a possibility to present and discuss the news 
accumulated during the year. Another important specific 
feature was always an opportunity for a large number of 
physicists from the former USSR and other East European countries 
for whom distant trips were difficult for the financial (earlier 
also for bureaucratic) reasons to participate to the conference. 

The special feature of this conference was a wider geography and 
a larger number of participants (113 persons and among them only 
12 women) from the countries they represented: Algeria -1, 
Belarus-4, Belgium -1, Bulgaria -2, China -2, Czech Republic -5, 
Estonia -1, France -5, Germany -4, Holland -1, India -2, Iran -1, 
Italy -2, Poland -5, Portugal -1, Russia -25, Slovakia -1, Sweden 
-1, Switzerland -1, the UK -1, Ukraine -2, the USA -11, 
Uzbekistan -1, Vietnam -1. As always, many physicists from JINR 
(about 35) participated to the conference. The reason for the 
increasing popularity of the conference became, apparently, the 
fact that this year brought many new experimental results. Some 
of them were for the first time presented in Dubna. 

X.Artru in his work, together with Z.~Belghobsi, proposed a simple 
explanation of the Collins effect and the effect of jet 
handedness in the model of sequential fragmentation of quark and 
proposed the program of realization of the model in the Monte 
Carlo method. Also preliminary results of the new measurements of 
the structure function g$_{2 }$ by the HERMES collaboration 
(A.Ivanilov) were reported for the first time. 

Classical experiments on the study of the nucleon spin structure 
at high energies use both scattering leptons on polarized 
nucleons (HERMES, JLab, COMPASS) and collisions of the polarized 
protons (RHIC, IHEP, JINR). The joint description of such 
different high-energy processes becomes possible due to the 
application of the fundamental theory of strong interactions, 
quantum chromodynamics (QCD), and remarkable properties of 
factorization, local quark-hadron duality and asymptotic freedom, 
which allow one to calculate the characteristics of a process 
within the framework of perturbation theory (PT). At the same 
time, parton distribution functions, correlation and 
fragmentation functions, which are not calculable and therefore 
require modeling methods, are universal and do not depend on the 
process. A number of reports at the conference were dedicated to 
the development and application of this type of models (P.Zavada 
- the original covariant model of nucleon, J.Soffer - quantum 
statistical model, N.Sharma - chiral model of constituent quarks 
and others). 

The theoretical description of processes with the participation 
of spin and especially an internal transverse parton motion 
proves to be, as always, more complicated, so that the number  of 
such functions increases and the picture connected with them 
loses to a considerable degree the simplicity of a parton model 
with its probabilistic interpretation. One of the difficulties 
here is the evolution of these functions with a change in the 
wavelength of a tester. An approach to its solution was presented 
in the talk of I.Cherednikov. 

The quark helicity distributions in a nucleon are the most well 
studied so far. The results of their more accurate measurements 
by the COMPASS (Y. Bedfer) and CLAS (Y.Prok) collaborations were 
presented. Contemporary experimental data are sufficiently 
precise to include in their QCD-analysis not only the correction 
of perturbation theory but also contributions of higher twists 
(A.Sidorov, O.Shevchenko, V.Khandramay, E.Christova, G.Ramsey, 
H.Dahiya, D.Str\'{o}zik-Kotlorz, F.Arbabifar). In this case, the 
positive polarization of strange quarks is excluded with high 
probability. However, the polarization of gluons agrees with the 
results of their direct measurement (although, with large 
uncertainty thus far) by the COMPASS (K. Kurek, C.Franco) and 
PHENIX+STAR (A.Bazilevsky, D.Svirida, I.Alekseev) collaborations, 
and its low value seems insufficient for resolving the so-called 
nucleon spin crisis. 

Hope for its overcoming is now on the contributions of the orbital angular 
momenta of quarks and gluons which can be determined by measuring the 
so-called Generalized Parton Distributions (GPD). The 15-year history of 
their appearance and current situation was dwelled upon in the talk by 
A.Radyushkin - one of the founders of this direction in QCD. Different 
theoretical aspects of GPDs were considered in the talks by S.Goloskokov, 
S.Manaenkov, L.Szimanovski and K.Semenov-Tyan-Shanskiy. Different 
experimental aspects of their measurements and preparation for new ones were 
presented in the talks of A.Sandacz, A.Morreale and P.Sznajder (COMPASS), 
V.Korotkov (HERMES) and V.Kubarovsky (JLab). 

Other important spin distribution functions manifest themselves in 
scattering of transversely polarized particles. The processes in which the 
polarization of only one particle (initial or final) is known are especially 
interesting and complicated from the theoretical point of view (and 
relatively simple from the point of view of experiment -- such 
complementarities frequently occur). Such single spin asymmetries are 
related to the T-odd effects, i.e. they seemingly break invariance under 
time reversal. Here, however, we deal with an effective breaking connected 
not with the true noninvariance of fundamental (in our case, strong, 
described by QCD) interaction under time reversal, but with its 
simulations by thin effects of rescattering in the final or initial state. 

The effects of single asymmetry have been studied by theorists 
(including Dubna theorists who have priority in a number of 
directions) for more than 20 years, but their study received a 
new impetus in recent years in connection with new experimental 
data on the single spin asymmetry in the semi-inclusive 
electro-production of hadrons off a longitudinally and 
transversely polarized targets (HERMES - V.Korotkov, CLAS - 
Y.Prok and COMPASS - C.Adolph, S.Elia).  In particular, HERMES 
data on the so-called "Sivers distribution function" for 
secondary pions, related to the left-right asymmetry of parton 
distribution in transversely polarized hadron, are described by 
the existing theory. However, the data for positive kaons in the 
region of small $x$ approximately 2,5 times larger than its 
predictions, which could testify to an essential role of a 
antiquark Sivers function. However, the new measurements of this 
asymmetry by the COMPASS collaboration do not confirm {such big} 
deviation, which favors of another possibility -- the influence 
of higher twist contributions.

New data on the single spin asymmetries of secondary pions and 
$\eta$-mesons in polarized proton-proton collisions with the 
energies RHIC (200x200 GeV) were presented by the PHENIX 
(O.Eyser) collaboration. They confirm amazingly large asymmetries 
in the region of the fragmentation of the polarized proton and 
their drop to zero in the central region of rapidities and the 
region of the nonpolarized proton beam obtained earlier at lower 
energies. This confirms their energy independence. However, 
PHENIX does not see a large difference in the asymmetries of 
$\eta $- and $\pi ^{0}$-mesons obtained earlier by STAR. At the 
same time particular mechanisms of the origin of these 
asymmetries remain a riddle so far. 

Thus, although single spin asymmetries on the whole are described 
by the existing theory, their development continues (I.Anikin). 
The T-odd distribution functions appearing here lose key 
properties of universality and become ``effective'', dependent on 
the process in which they are observed. In particular, the most 
fundamental QCD prediction is the change of the sign of the 
Sivers function in passing from the pion electro-production 
process to the Drell-Yan pair production on a transversely 
polarized target. This conclusion is planned to be checked in the 
COMPASS experiment (A.Guskov) and at colliders RHIC (L.Nogach), 
NICA and PANDA-PAX (M.Destefanis). 

Significant interest and discussions were caused by new JLab data 
presented at the conference on measurements of the ratio of 
proton electric and magnetic form factors performed by the 
"technique of the recoil polarization" (Ch.Perdrisat). The 
previous JLab measurements showed that this relation was not 
constant, as it was considered for a long time, but linearly 
decreases with increase of momentum transfer $Q^2$ (the so-called 
"form factor crisis"). New data obtained in the past year 
(experiment GEp(III) with JINR participation) indicate flattening 
of this ratio in the Q$^{2}$=6-8 [GeV]$^{2}$ region. A question 
whether this behavior is due to incomplete calculation of 
radiative corrections, in particular, two-photon exchange, 
remains open yet. 

As always, the sources of polarized particles (M.Chetvertkov, Yu.Plis, 
D.Karlovets), physics of the acceleration of polarized beams 
(Yu.Kondratenko), physics of polarimeters (V.Ladygin, A.Zelenskiy, M.Runtso, 
D.Smirnov), and the polarized target technique (Yu.Kiselev) were discussed 
at the conference. 

Great interest was generated by the first results of experiments at 
the Large Hadron Collider (LHC) in CERN relating to spin physics 
(C. Buszello). In particular, the determination of spin and 
quantum numbers of Higgs- and Z-bosons, polarization of W, and 
also the spin phenomena in heavy quark physics. A number of talks 
were devoted to theoretical possibilities of Z' search and other 
exotics at LHC and future International Linear Collider (ILC) of 
electrons (V.Andreev, A.Tsitrinov, J.K\"{o}rner).

Finally, considerable attention was given to the projects of 
further development of polarization studies. A large and detailed 
report about the project of eRHIC (collider of polarized protons 
of 250 GeV and nuclei with polarized electrons of 20 GeV) at BNL 
made by E.Aschenauer. Having a large luminosity (10$^{34})$, it 
will make it possible to increase the accuracy in measurement of 
gluon and quark spin distribution functions in a proton, as well 
as GPDs by an order of magnitude. The plans of further studies at 
the modified accelerator at JLab (Y.Prok, V.Kubarovsky) were also 
discussed. The program of polarized proton beam formation from 
decay of Lambda-particles at the IHEP accelerator U-70 in 
Protvino for spin studies at the installation SPASCHARM under 
construction was presented by S.Nurushev. He emphasized the 
importance of the comparative study of the spin effects induced 
by particles and antiparticles. However special interest was 
caused by the plans of creation at IKP (J\"{u}ulich) of a unique 
European complex for determining the electric dipole moment (EDM) 
of a proton and nuclei (very detailed talk of N.Nikolaev). The 
matter is that the dipole moment of fundamental particles 
violates both space and time parity and its detection would 
indicate violation of the Standard Model and, in particular, a 
possibility of approach to the problem of understanding of baryon 
asymmetry of the Universe. The projected complex will make it 
possible to lower the limit of deuteron EDM measurement up to 
10$^{ - 24}$. 

The reports relate on the development at LHEP accelerating 
complex of JINR were also presented in the program of the 
conference (R.Kurilkin, N.Ladygina). The newest methods and the 
results of calculations of specific features of spin dynamics 
under acceleration at the Nuclotron of polarized protons and the 
lightest nuclei were also reflected (Yu.Kondratenko). Some new 
proposals for conducting polarization studies on the basis of the 
modernized complex Nuclotron-M and at the complex NICA projected 
at JINR were presented (O.Teryaev, O.Selyugin). Within the 
framework of DSPIN-11 two working discussions (leader 
A.Kovalenko) of vital problems of the infrastructure development 
for further studies in spin physics at the complex Nuclotron/NICA 
took place in which specialists of JINR, BNL, MEPI, ITEP and INR 
participated. Participants heard information about the project 
``SPRINT'' (Spin Physics Research of Infrastructure at Nuclotron) 
being developed at LHEP, about polarimetry at the complex 
AGS/RHIC at BNL, in particular, problems of development and use 
of CNI-polarimeters and possibility of their use at the NICA 
collider, and other questions. 

The spin community represented at the conference supported these 
plans to create new unique possibilities for conducting 
polarization studies at the accelerating complex of LHEP at JINR. 
The accelerating complex with such potentialities will not have 
competitions from other centers carrying out polarization 
studies, and the obtained data will help to solve the riddles of 
the spin effects which have not been solved since the 70s of the 
past century. 

The summary of the meeting was made in the final report by J.Soffer. 

The success of the conference was due to the support by the 
Russian Foundation for Basic Research, International Committee 
for Spin Physics, ``Dynasty'' Foundation, European Physical 
Society and the JINR programs of for international collaboration: 
Heisenberg-Landau, Bogoliyubov-Infeld and Blokhintsev-Votruba 
ones. This made it possible to provide noticeable financial 
support to participants from Russia and other JINR Member States. 
The materials of the conference, including slides of all 
presented talks, and Proceedings (412 pages) are available on the 
site: {\tt http://theor.jinr.ru/$\sim$spin/2011/}.

\end{document}